\documentclass{nst}

\usepackage{subfigure,dcolumn}
\usepackage{epstopdf}
\usepackage{mhchem}
\usepackage{upgreek}
\usepackage{relsize}
\usepackage{xcolor}

\usepackage{booktabs}    
\usepackage{multirow}     
\usepackage{tabularx}     
\usepackage{tikz}
\usepackage{xspace}

\newcommand{\pAu}           {$p + \mathrm{Au}$\xspace}
\newcommand{\SC}            {$SC$\xspace}
\newcommand{\pA}            {$p + A$\xspace}
\newcommand{\AuAu}          {$\mathrm{Au} + \mathrm{Au}$\xspace}

\newcommand{\Au}            {$\mathrm{Au}$\xspace}
\newcommand{\AMPT}          {$\texttt{AMPT}$\xspace}
\newcommand{\UrQMD}         {$\texttt{UrQMD}$\xspace}
\newcommand{\HIJING}        {$\texttt{HIJING}$\xspace}
\newcommand{\SM}            {$\texttt{SM}$\xspace}
\newcommand{\ZPC}           {$\texttt{ZPC}$\xspace}
\newcommand{\QCD}           {$\texttt{QCD}$\xspace}
\newcommand{\LQCD}          {$\texttt{LQCD}$\xspace}
\newcommand{\pQCD}          {$\texttt{pQCD}$\xspace}

\newcommand{\POT}           {$\texttt{POT}$\xspace}
\newcommand{\Dy}            {\Delta y}
\newcommand{\sNN}           {\sqrt{s_{_{\rm NN}}}}
\newcommand{\sameHTKpT}     {P^{\text{same}}_{H_T K^+_T}}
\newcommand{\sameHPKpP}     {P^{\text{same}}_{H_P K^+_P}}
\newcommand{\mixHTKpT}      {P^{\text{mix}}_{H_T K^+_T}}

\newcommand{\sameHKp}       {P^{\text{same}}_{HK^+}}
\newcommand{\mixHbKp}       {P^{\text{mix}}_{\overline{H}K^+}}
\newcommand{\sameHbKm}      {P^{\text{same}}_{\overline{H}K^-}}
\newcommand{\sameHbPKmP}    {P^{\text{same}}_{\overline{H}_PK^-_P}}
\newcommand{\sameHKm}       {P^{\text{same}}_{HK^-}}
\newcommand{\sameHbKp}      {P^{\text{same}}_{\overline{H}K^+}}
\newcommand{\mixHbKm}       {P^{\text{mix}}_{\overline{H} K^-}}
\newcommand{\mixHKp}        {P^{\text{mix}}_{H K^+}}
\newcommand{\mixHKm}        {P^{\text{mix}}_{H K^-}}
\newcommand{\CBS}           {C^{\text{CBS}}_{H K^+}}
\newcommand{\CBSyp}         {C^{\text{CBS}}_{H_{y>0} K^+}}
\newcommand{\CBSym}         {C^{\text{CBS}}_{H_{y<0} K^+}}
\newcommand{\CBSLambda}     {C^{\text{CBS}}_{\Lambda K^+}}
\newcommand{\CBSLambdayp}   {C^{\text{CBS}}_{\Lambda_{y>0} K^+}}
\newcommand{\CBSLambdaym}   {C^{\text{CBS}}_{\Lambda_{y<0} K^+}}
\newcommand{\CBSXi}         {C^{\text{CBS}}_{\Xi K^+}}
\newcommand{\CBSXiyp}       {C^{\text{CBS}}_{\Xi_{y>0} K^+}}
\newcommand{\CBSXiym}       {C^{\text{CBS}}_{\Xi_{y<0} K^+}}
\newcommand{\EMDp}          {\texttt{EMDp}}
\newcommand{\EMD}           {\texttt{EMD}}

\begin{document}

\title{Studying baryon number transport dynamics via hyperon-kaon correlations in \pAu collisions at $\sNN=20$, $39$ and $62$ GeV}

\author{Siyuan Ping}
\email{syping22@m.fudan.edu.cn}
\affiliation{Key Laboratory of Nuclear Physics and Ion-beam  Application (MOE), Fudan University, Shanghai 200433, China}
\author{Xiaozhou Yu}
\affiliation{Key Laboratory of Nuclear Physics and Ion-beam  Application (MOE), Fudan University, Shanghai 200433, China}
\author{Long Ma}
\affiliation{Key Laboratory of Nuclear Physics and Ion-beam  Application (MOE), Fudan University, Shanghai 200433, China}

\begin{abstract}
The observation of positive net hyperon baryon numbers at mid-rapidity in heavy-ion collisions indicates that baryon numbers from incident nucleons can be transported across a large rapidity gap to hyperons where strange quarks of $s-\bar{s}$ are pair-produced. Consequently, hyperons and kaons are expected to be correlated, providing a sensitive probe of both baryon number transport mechanism and strange quark pair correlation. Such correlation may be used to test the gluon-junction interaction mechanism where a $Y$-shaped gluonic field may carry the baryon number and be responsible for the baryon number transport to hyperons over a large rapidity gap. We present hyperon-kaon correlations as a function of their relative rapidity in \pAu collisions at $\sqrt{s_{NN}} = 20$, $39$, and $62$ GeV using a multiphase transport (\AMPT) model and Ultra-relativistic Quantum Molecular Dynamics (\UrQMD) models where hyperon-kaon pairs are originated from fragmentation scheme. We quantify the correlation function using the Wasserstein distance method and systematically investigate the correlation dependence on the beam energy, hyperon emission direction and detector acceptance. Our simulation results provide a baseline without the baryon junction mechanism for future experimental measurements.
\end{abstract}


\maketitle

\section{Introduction}\label{Introduction}
A baryon-to–antibaryon ratio greater than unity has been observed at mid-rapidity ($y \sim 0$) in heavy-ion collisions at Relativistic Heavy Ion Collider (RHIC) and Large Hadron Collider (LHC) energies~\cite{beardenNuclearStoppingCollisions2004, collaborationStrangeHadronProduction2020, AnticicTPRL2004, AbelevBIPRC2007, AnticicTPRL2005, AnticicTPRC2008, AnticicTPRC2009, AntinoriPLB2004, AntinoriJPG2006, AdamsPRL2007}.
Since the initial baryons are located at beam rapidity $y_{\text{beam}}$ and baryon number is strictly conserved in the Standard Model, this observation implies that a fraction of the baryon number must be transported from beam rapidity to mid-rapidity across a large rapidity gap $\delta y \sim y_{\text{beam}}$.
This phenomenon is commonly referred to as baryon number transport (BNT) \cite{LewisLvEPJC2024, LvLiCPC2024}, and can be described by an framework named gluon junction (or baryon junction) model that construct a Y-shaped gluonic topology that connects to the three valence quarks, carries a unit baryon number and constitutes the only gauge-invariant state configuration for a baryon in quantum chromodynamics (\QCD)~\cite{rossiPossibleDescriptionBaryon1977, kharzeevCanGluonsTrace1996, rossiAspectsofBaryoniumPhysics1977}. This Y-Ansatz has been verified by Lattice \QCD (\LQCD) \cite{TakahashiPRL2001}.
In contrast to the valence-quark picture, where the baryon number is carried solely by three valence quarks, the gluon junction model naturally leads to more significant BNT. This is because the gluonic field (valence quarks) in a baryon typically carries a smaller (larger) fraction of the baryon momentum, resulting in a larger (smaller) interaction cross section and more (less) effective stopping toward mid-rapidity.

\begin{figure*}[!htb]
  \centering

  \includegraphics[width=0.9\textwidth]{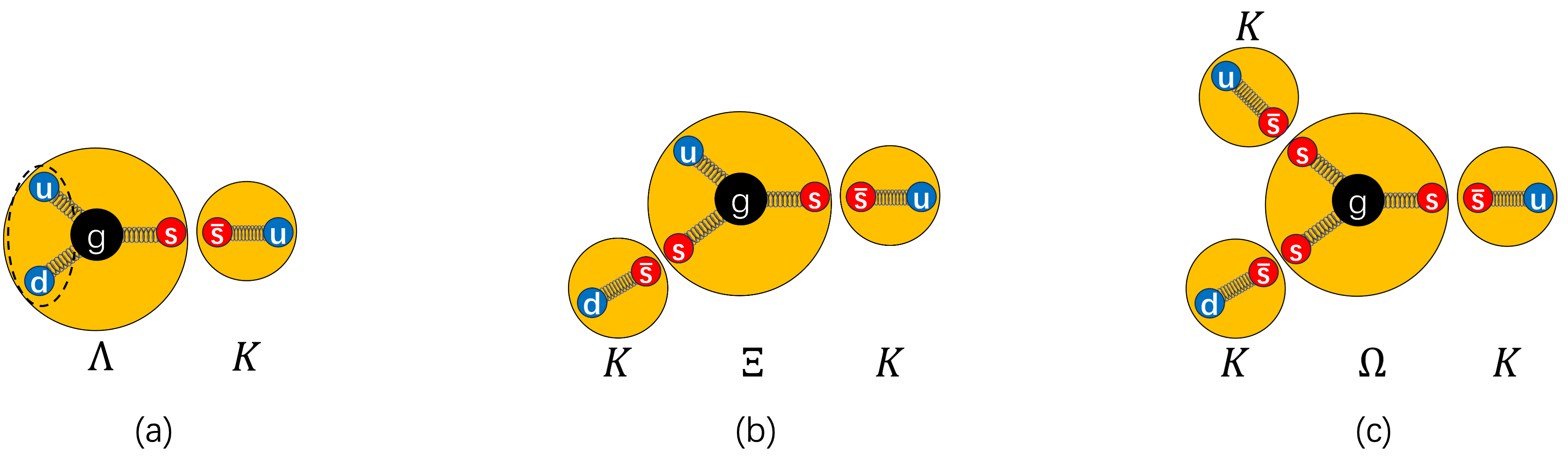}

  \caption{Schematic illustration of proton fragmentation into (a) $\Lambda$ with one kaon, (b) $\Xi$ with two kaons, and (c) $\Omega$ with three kaons, corresponding to the breaking of one, two, and three $s-\bar{s}$ quark pairs, respectively, within the gluon junction mechanism. (a): diquark--quark fragmentation; (b): three-valence-quark fragmentation; (c): baryon junction fragmentation, where the gluon junction is completely separated from all valence quarks.}
  \label{JunctionDisplay}
\end{figure*}

Recently, in \AuAu collisions, correlations among strange hadrons have been proposed as a potential probe of baryon number transport (BNT) in momentum and rapidity phase space~\cite{dongStudyBaryonNumber2024}.
Since the initial nucleons carry no net strangeness, the conservation of strangeness (\SC) in strong interactions implies that $s$–$\overline{s}$ quark pairs produced from the vacuum can naturally generate correlations between hyperons and kaons.
These hyperon–kaon dynamical correlations are influenced by the specific fragmentation process, as illustrated in Fig.~\ref{JunctionDisplay}, and can therefore serve as a sensitive probe for testing baryon fragmentation mechanisms. 

In addition, Proton–nucleus (\pA) collisions may provide a more favorable environment for constraining theoretical models, since baryons emitted into the projectile (proton-going) hemisphere are more likely to carry baryon number originating from the incident proton, thereby exhibiting a clearer fragmentation pattern of the projectile proton~\cite{HuangHeavyIonPhysicsfromBevalactoRHIC, collaborationProtonFragmentationBaryon2002}. 
Baryons that carry baryon number transported from the incident proton are expected to undergo, on average, a larger number of collisions, whereas those originating from the \Au nucleus may experience a different collision history. Consequently, hyperons emitted in the proton-going and Au-going directions are expected to correlate with kaons differently. 

In this paper, we introduce new observables to measure the hyperon-kaon correlation in Secs.~\ref{Sec_CBS} and~\ref{Sec_EMDp}, based on a combinatorial background subtraction method, which removes combinatorial backgrounds arising from the mixing of strange hadrons originating from different sources. The correlation results from the proton-going and Au-going cases are compared to study the sensitivity for observing the BNT effect in \pA collisions.
 
\section{Methodology}\label{Methodology}

\subsection{Hyperons Production and Strangeness Conservation}\label{sec.2.2}
\begin{figure}
    \includegraphics[width=0.65\hsize]{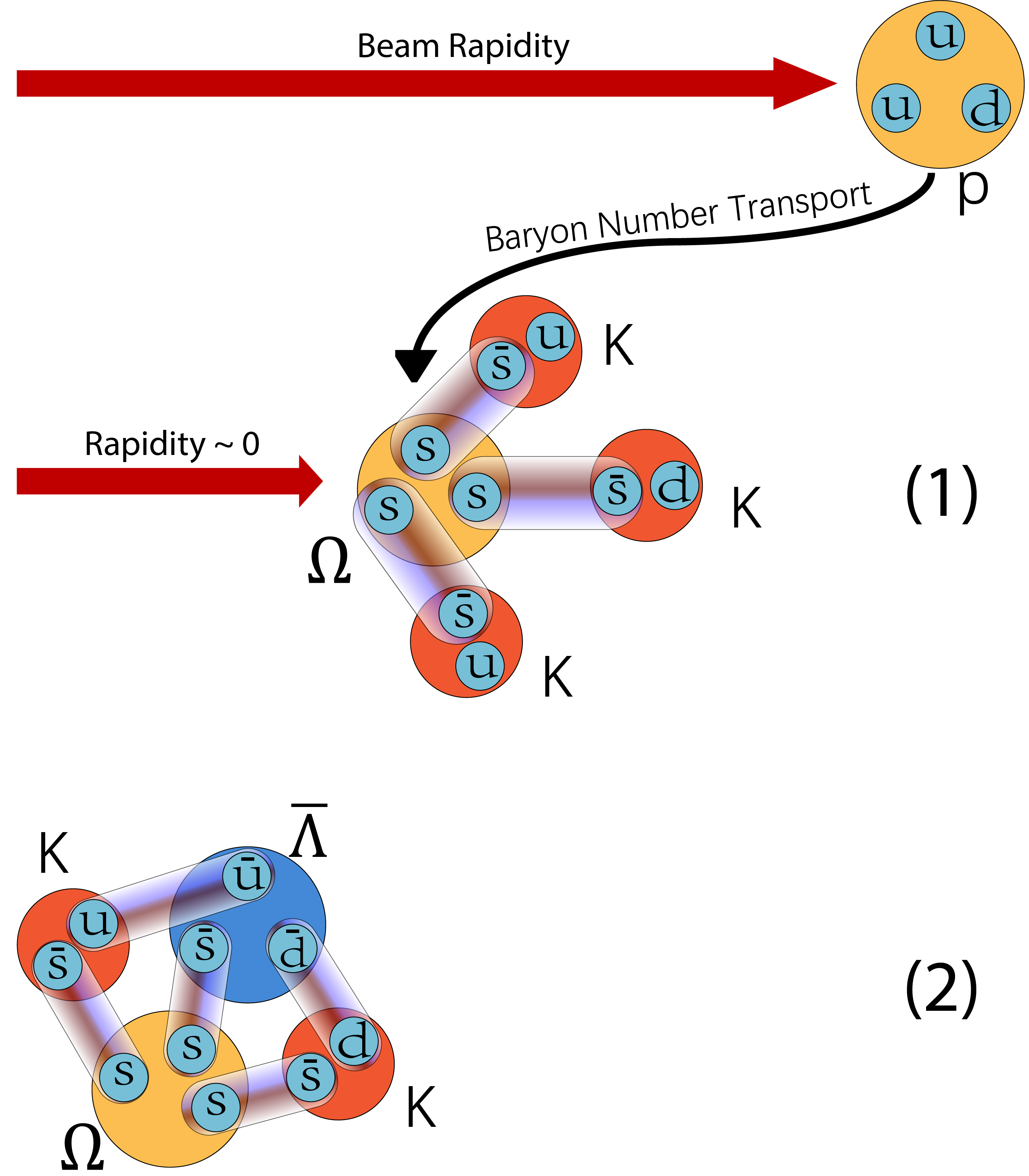}
    \caption{Schematic illustration of two production mechanisms. 
    In scenario (1), an incident proton transports its baryon number to an $\Omega$ hyperon via general associated production. 
    In scenario (2), one $\Omega$, one $\bar{\Lambda}$, and two kaons are produced from the vacuum through general pair production. }    
    \label{SchemesCartoon}
\end{figure}

The valence strangeness quarks in produced hyperons originate from $s$–$\overline{s}$ pairs production following two scenarios: (1) general associated production and (2) general pair production\cite{dongStudyBaryonNumber2024}. 
For example, in Figure~\ref{SchemesCartoon}, two $\Omega$ production schemes are displayed. 
In general associated production, three $s$-$\overline{s}$ pairs are produced from vacuum due to the strangeness conservation. Three $s$ quarks coalesced an $\Omega$, and three $\overline{s}$ along with the valence quarks ($uud$) from a proton form three kaons. 
Therefore, this new generated $\Omega$ carry the baryon number transported from a proton. 
In general pair production, all valence quarks are produced from vacuum, so no baryon number is transported. 
The initial protons with beam rapidity transport baryon number to $\Omega$ in middle rapidity region, showing baryon stopping effect. 

The $s-\overline{s}$ pair production indicates a correlation between hyperons and kaons in both scenarios shown in Figure~\ref{SchemesCartoon}, suggesting a possibility to study the baryon number transport by measuring the correlation between hyperons and kaons. 
In Table~\ref{SchemesRatio}, one can see the hyperon correlated number of kaons in different scenarios. 
Since scenario 2 does not involve BNT effect while still generating hyperon–kaon correlations, its contribution must be removed in order to disentangle the genuine BNT signal. The subtraction procedure is described in Sec.~\ref{sec.2.3}. 

The production of lighter hyperons, such as $\Lambda$ (uds) and $\Xi$ (uss), follows a similar mechanism. However, their $u$ and $d$ valence quarks may originate from the incident baryons. In contrast, for incident baryons fragmenting into $\Omega$ (sss) hyperons, all valence quarks must be transferred to mesons. Therefore, in scenario 1, the dynamical difference between $\Omega$ hyperons and kaons directly reflects the separation between gluon-field and valence-quark.
In this sense, the $\Omega$ baryon may provide the cleanest configuration among the three hyperon species for studying hyperon–kaon correlations.


\begin{table}[!htb]
    \caption{Summary of two scenarios of hyperon production.}
    \label{SchemesRatio}
    \begin{tabularx}{8cm}{@{\extracolsep{\fill}} c c c c c}
    \toprule
    Scenario & Hyperon & Kaons number & SC & BNT \\
    \midrule
    1 & $\Lambda$ & 1            & $\sqrt{}$ & $\sqrt{}$ \\
    1 & $\Xi$     & 2            & $\sqrt{}$ & $\sqrt{}$ \\
    1 & $\Omega$  & 3            & $\sqrt{}$ & $\sqrt{}$ \\
    2 & $\Lambda$ & 0 or 1       & $\sqrt{}$ &           \\
    2 & $\Xi$     & 0, 1 or 2    & $\sqrt{}$ &           \\
    2 & $\Omega$  & 0, 1, 2 or 3 & $\sqrt{}$ &           \\
    \bottomrule
    \end{tabularx}
\end{table}

\subsection{Model Description}\label{ModelDescription}

In this work, we use A multi-phase transport (\AMPT) model \cite{linFurtherDevelopmentsMultiphase2021, linMultiphaseTransportModel2005} with string melting (\SM) mechanism \cite{linPartonicEffectsElliptic2002, linPartonicEffectsPion2002} and the Ultra-relativistic Quantum Molecular Dynamics (\UrQMD) model \cite{bassMicroscopicModelsUltrarelativistic1998, bleicherRelativisticHadronhadronCollisions1999} to simulate \pAu collisions at $\sNN=20$, $39$ and $62$ GeV. More than 100 million events are generated for both simulation datasets. 

\subsubsection{\AMPT}
The \AMPT model has been applied extensively and successfully to study heavy-ion collisions at relativistic energies\cite{JYZhang2025, BSXi2025, LMa2021}. In the model simulation, the initial minijet partons and soft excited strings are generated by the \HIJING event generator~\cite{wangRoleMultipleMinijets1991, wangHijingMonteCarlo1991, wangSystematicStudyParticle1992, wangHIJING10Monte1994}, providing both spatial and momentum information.
In the string-melting (\SM) version of AMPT, the excited strings associated with participant nucleons are converted into partons, together with the minijet partons. The partonic evolution is then handled by the Zhang’s Parton Cascade (\ZPC)~\cite{zhangZPC101Parton1998} program, which solves the Boltzmann equation for two-body partonic scatterings, where the parton–parton scattering cross sections are calculated within perturbative quantum chromodynamics (\pQCD) scheme.
The high-density partons output from \ZPC subsequently undergoes hadronization via a quark coalescence mechanism, in which the nearest two or three partons in phase space recombine to form hadrons in the \SM version.
After hadronization, hadron cascade is simulated using the relativistic hadronic transport (ART) model~\cite{liFormationSuperdenseHadronic1995}, which includes both elastic and inelastic hadron–hadron scatterings. 



\subsubsection{\UrQMD}
\UrQMD ~\cite{bassMicroscopicModelsUltrarelativistic1998, bleicherRelativisticHadronhadronCollisions1999} is a hadronic relativistic transport model that initialize the projectile and target nuclei according to a Gaussian-like wave packet. The nucleon–nucleon interactions are discribed by solving non-relativistic and density-dependent Hamiltonian equations of motion, supplemented by Yukawa and Coulomb potentials. 
The collision term in \UrQMD is implemented as a sequence of binary interactions determined by geometric collision criteria and experimentally constrained hadronic cross sections, including elastic and inelastic scatterings, resonance formation and decay. 
At sufficiently high energies, particle production in UrQMD is described by string excitation and subsequent fragmentation, providing an effective mechanism for multi-particle production that complements resonance-dominated processes at lower energies. The \UrQMD version 4.0 is used in our simulation. 



\subsection{Hyperon-Kaon pair distributions}\label{sec.2.3}

We calculate the $\Lambda-K$ and $\Xi-K$ correlation through their pair distributions in relative rapidity difference between one kaon and one hyperon in each pair:
\begin{align}\label{08032107}
    \Dy = \theta(y_{H})(y_K-y_{H}) + \theta(-y_{H})(y_H-y_K),
\end{align}
where the step function $\theta(x)$ is defined as
\begin{align}
    \theta(x)=
    \begin{cases}
        1, & x\geq 0 \\
        0, & x < 0
    \end{cases}.
\end{align}
Here $y_K$ and $y_H$ represent the rapidities of the kaon and hyperon respectively. 
In this study, we define the initial proton as flying in the same direction of z axis and Au as flying in the opposite direction of z axis. Thus, initial proton (Au) carries positive (negative) rapidity, respectively. 
A positive (negative) $\Dy$ value indicates that the kaon has larger (smaller) rapidity than the hyperon, regardless of the hyperon's emission direction.

In this work, we study the correlations without imposing any acceptance cuts on particles. For comparison with experimental results, we also apply a pseudo-rapidity acceptance cut of $\lvert \eta \rvert < 1.5$ to all particles, considering typical detector designs~\cite{shaoExtensiveParticleIdentification2006, collaborationPhysicsProgramSTAR2016}.
This pseudo-rapidity range approximately corresponds to the mid-rapidity region for hyperons and kaons, where hyperons experience stronger baryon stopping effects.
According to Eq.~(\ref{08032107}), this acceptance constraint restricts the range of $\Dy$ to (-3.0,1.5). 

Hyperons can be produced either through general associated production (denoted as $H_T$) or through general pair production (denoted as $H_P$), whereas anti-hyperons can only be produced via general pair production (denoted as $\overline{H}_P$).
The $K^+$ mesons may originate from general associated production, general pair production, or processes without any interaction with hyperons (represented as $K^+_T$, $K^+_P$, and $K^+_U$, respectively).
Similarly, $K^-$ mesons can be categorized as $K^-_P$ and $K^-_U$, since $K^-$ cannot be produced in processes involving baryon number transport.

We employed the event-mixing technique, similar to ~\cite{drijardStudyEventMixing1984} in this study. Hyperon-kaon pairs on an event-by-event basis are used to construct the same-event distribution, while pairs from different events are used to build the mixed-event distribution. 
We focus on:
\begin{widetext}
    \begin{align}
        \sameHKp(\Dy) &= \sameHTKpT(\Dy) + \sameHPKpP(\Dy) + P_{H_TK^+_P}(\Dy) + P_{H_PK^+_T}(\Dy) + P_{H_TK^+_U}(\Dy) + P_{H_PK^+_U}(\Dy) \, , \label{2509072052} \\
        \mixHbKp(\Dy) &= P_{\overline{H}_PK^+_P}(\Dy) + P_{\overline{H}_PK^+_T}(\Dy) + P_{\overline{H}_PK^+_U}(\Dy) \, , \label{2509072054} \\
        \sameHbKm(\Dy) &= \sameHbPKmP(\Dy) + P_{\overline{H}_PK^-_U}(\Dy) \, , \\
        P^{\text{mix}}_{HK^-}(\Dy) &= P_{H_PK^-_P}(\Dy) + P_{H_TK^-_P}(\Dy) + P_{H_TK^-_U}(\Dy) + P_{H_PK^-_U}(\Dy) \, . \label{2509261045}
    \end{align}
\end{widetext}
Here, in Eq.~(\ref{2509072052})-(\ref{2509261045}), the superscripts “same” and “mix” denote pair distributions constructed from same-event and mixed-event, respectively, with the latter serving as the uncorrelated reference. 
Each mixed-event pair distribution is normalized to have the same total number of pairs as its corresponding same-event distribution
\begin{align}
    \sum_{\text{pairs}}\mixHKp(\Dy)&=\sum_{\text{pairs}}\sameHKp(\Dy) \, , \label{2601021531}\\
    \sum_{\text{pairs}}P^{\text{mix}}_{HK^-}(\Dy)&=\sum_{\text{pairs}}\sameHKm(\Dy) \, , \\
    \sum_{\text{pairs}}\mixHbKm(\Dy)&=\sum_{\text{pairs}}\sameHbKm(\Dy) \, , \\
    \sum_{\text{pairs}}\mixHbKp(\Dy)&=\sum_{\text{pairs}}\sameHbKp(\Dy) \, . \label{2601021532}
\end{align}
The pair distributions without an explicit superscript imply that the same-event and mixed-event constructions are equivalent in those cases. 
The hyperons and anti-hyperons produced in scenario 2 are assumed to have equal probabilities of being produced, which implies that the rapidity distributions satisfy:
\begin{align}
    P_{K^+_P}(y) &= P_{K^-_P}(y) \, , \label{2509080907} \\ 
    P_{\overline{H}}(y) &= P_{H_P}(y) = P_{\overline{H}_P}(y) \, . \label{2509080908}
\end{align}
Also, it is reasonable to assume that
\begin{align}
    P_{K^+_U}(y) = P_{K^-_U}(y) \, , \label{2509072053}
\end{align}
since the charge-conjugated kaons produced outside scenarios 1 and 2 should be generated symmetrically with respect to charge sign. 
From Eq.~(\ref{2509080907})-(\ref{2509072053}), one can get the pairs
\begin{align}
    \sameHPKpP(\Dy)&=\sameHbPKmP(\Dy) \, , \\ 
    P_{H_PK^+_U}(\Dy)&=P_{\overline{H}_PK^-_U}(\Dy) \, , \\
    P_{H_PK^-_U}(\Dy)&=P_{\overline{H}_PK^+_U}(\Dy) \, , \label{2409081710}\\
    P_{H_PK^-_P}(\Dy)&=P_{\overline{H}_PK^+_P}(\Dy) \, , \\
    P_{H_TK^+_P}(\Dy)&=P_{H_TK^-_P}(\Dy) \, , \\
    P_{H_TK^+_U}(\Dy)&=P_{H_TK^-_U}(\Dy) \, , \\
    P_{H_PK^+_T}(\Dy)&=P_{\overline{H}_PK^+_T}(\Dy)\,, \label{2409080916} \\
    P_{\overline{H}_PK^+_P}(\Dy)&=P_{\overline{H}_PK^-_P}(\Dy) \, . \label{2409081711}
\end{align}
Then, we derive the correlated kaon-hyperon pairs originating exclusively from general associated production:
\begin{widetext}
    \begin{align}
        \sameHTKpT(\Dy) = \sameHKp(\Dy) - \sameHbKm(\Dy) - P^{\text{mix}}_{HK^-}(\Dy) - \mixHbKp(\Dy) + 2\,\mixHbKm(\Dy) \, . \label{2509072207}
    \end{align}
\end{widetext}
Therefore we have removed the correlation contributed from the kaon-hyperon pairs produced in general pair production scenario or independently of hyperon production. 
Similarly, the uncorrelated pairs $\mixHTKpT(\Dy)$ can be obtained as
\begin{widetext}
    \begin{align}
        \mixHTKpT(\Dy) = \mixHKp(\Dy) + \mixHbKm(\Dy) - P^{\text{mix}}_{HK^-}(\Dy) - \mixHbKp(\Dy) \, . \label{2509172214}
    \end{align}
\end{widetext}
According to Eq.~(\ref{2601021531})-(\ref{2601021532}), $\mixHTKpT(\Dy)$ is also normalized to the same area with $\sameHTKpT(\Dy)$, that is
\begin{align}
    \sum_{\text{pairs}} \sameHTKpT(\Dy) = \sum_{\text{pairs}} \mixHTKpT(\Dy) \, .
\end{align}

\begin{figure}
    \includegraphics[width=0.65\hsize]{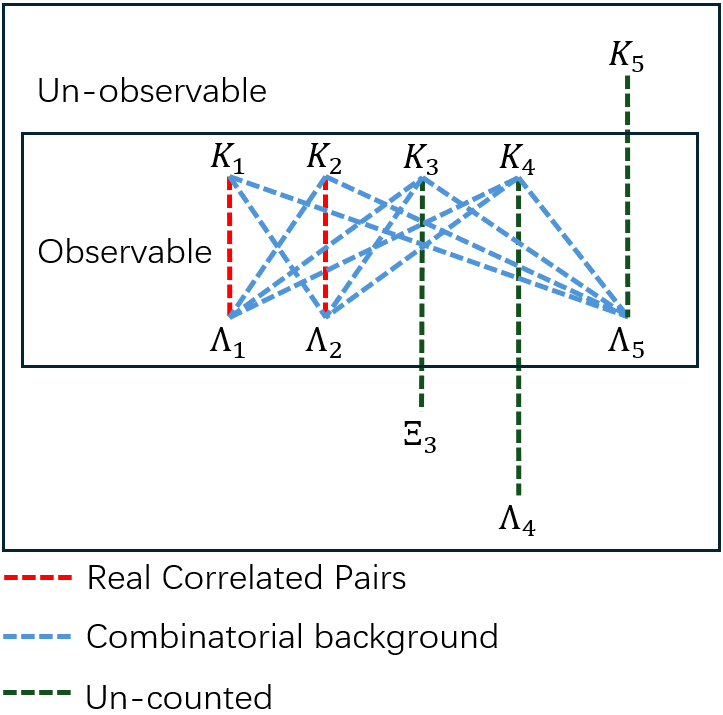}
    \caption{Schematic decomposition of the same-event distribution $P^{\text{same}}_{\Lambda_TK_T^+}(\Dy)$. In this example, $K_n$ is correlated with $\Lambda_n$ for $n=1,2$, while $K_3$ is correlated with $\Xi_3$. Due to detector acceptance limitations, $\Lambda_4$ and $K_5$ are not observable. Consequently, the distribution contains two correlated pairs and ten uncorrelated pairs.}
    \label{CombinatorialBackgroundDisplay}
\end{figure}

Although the hyperons and kaons in $\sameHTKpT(\Dy)$ originate from general associated production, some $K^+_T$ may still pair with irrelevant $H_T$ hyperons, forming a combinatorial background. 
Consequently, $\sameHTKpT(\Dy)$ contains both correlated and uncorrelated contributions. 
Figure~\ref{CombinatorialBackgroundDisplay} presents a schematic illustration of the $\Lambda_T$–$K^+_T$ pair distributions, where red lines indicate correlated $\Lambda$–$K^+$ pairs and blue lines indicate uncorrelated pairs (combinatorial background). 
The genuine correlation signal can be isolated by comparing same-event distributions with the corresponding mixed-event reference distributions. 

\subsection{Combinatorial background subtraction}\label{Sec_CBS}

\begin{figure}
    \subfigure{
        \label{Same_Mix_Sub_a}
        \includegraphics[width=0.65\hsize]{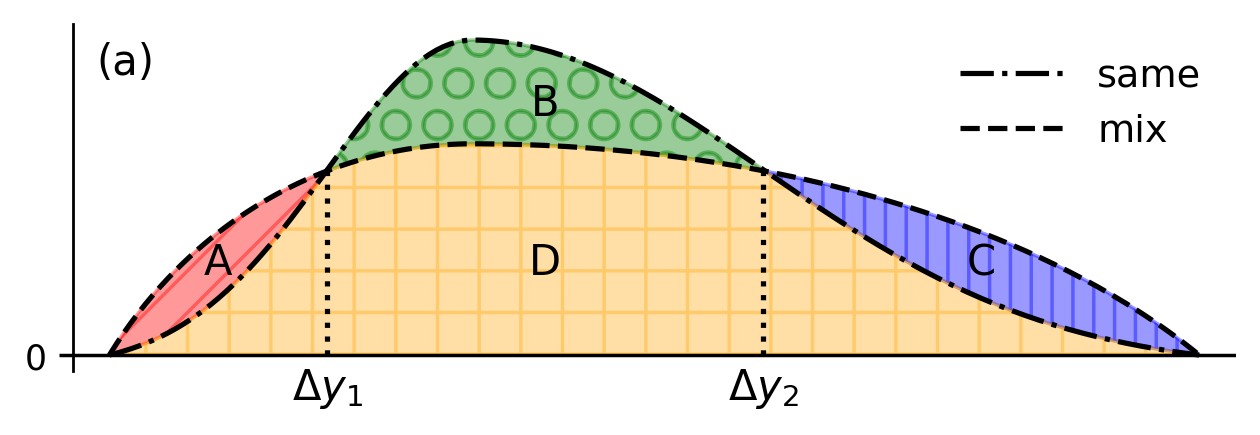}
    }
    \subfigure{
        \label{Same_Mix_Sub_b}
        \includegraphics[width=0.65\hsize]{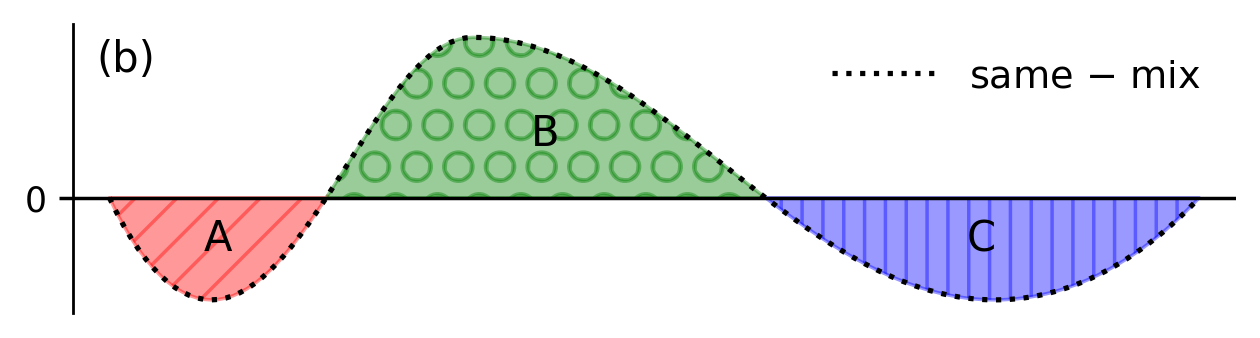}
    }
    \subfigure{
        \label{Same_Mix_Sub_c}
        \includegraphics[width=0.65\hsize]{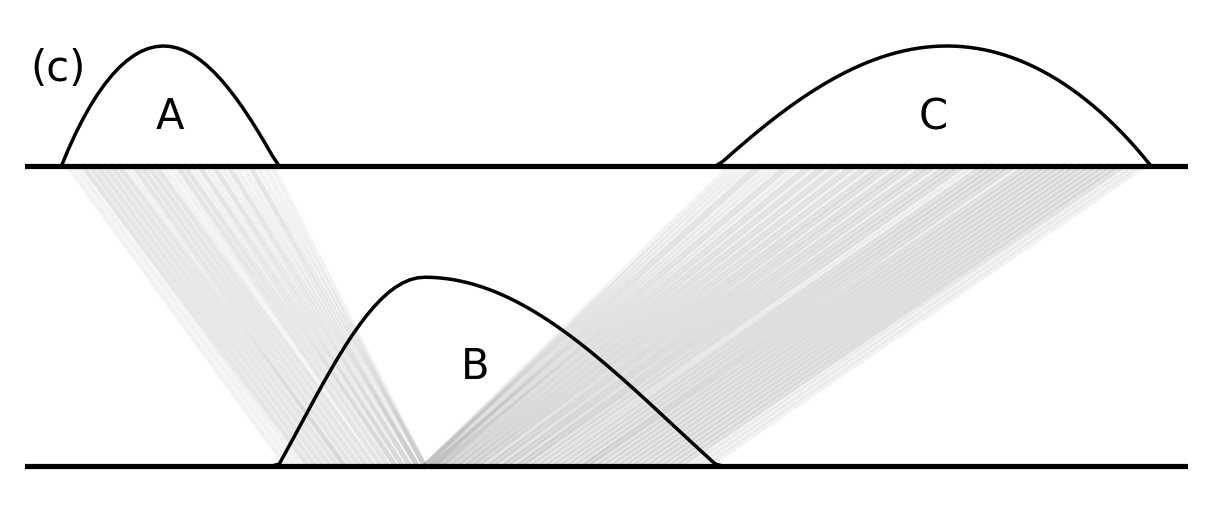}
    }
    \caption{Schematic illustration of (a) the same-event distribution (dash-dotted line) and the mixed-event distribution (dashed line); (b) the difference between the same-event and mixed-event distributions (dotted line); and (c) the optimal transport map that morphs area $A$ and $C$ into $B$. In region $B$ ($A$ and $C$), the same-event distribution is larger (smaller) than the mixed-event distribution.}
    \label{Same_Mix_Sub}
\end{figure}

Here we use the schematic plots in Fig.~\ref{Same_Mix_Sub} as an example to explain the hyperon–kaon correlation function which can be defined via combinatorial background subtraction as
\begin{align}
    \CBS(\Dy) = c\left[\sameHTKpT(\Dy) - \mixHTKpT(\Dy)\right]  \, , \label{2509191015}
\end{align}
where the normalization factor $c$ is chosen to satisfy
\begin{align}
    \sum_{\text{bins}} \Theta\left(\CBS(\Dy)\right) = \sum_{\text{bins}} \Theta\left(-\CBS(\Dy)\right)=1 \, ,
\end{align}
where $\Theta \left(x\right)$ is defined as
\begin{align}
    \Theta \left(x\right) = x\cdot \theta\left(x\right)\, .
\end{align}
The total positive ($\Theta\left(\CBS(\Dy)\right)$) and negative ($\Theta\left(-\CBS(\Dy)\right)$) areas are normalized to unity, corresponding to areas $B$ and $A+C$ in Fig.~\ref{Same_Mix_Sub}, respectively.
As discussed in Sec.~\ref{sec.2.3}, the same-event and mixed-event pair distributions can be decomposed into contributions from correlated pairs and combinatorial background:
\begin{align}
\sameHTKpT(\Dy) &= S(\Dy) + B'(\Dy) \, , \label{2509171532} \\
\mixHTKpT(\Dy) &= B(\Dy) + B'(\Dy) \, , \label{2509171533}
\end{align}
where $S(\Dy)$ represents the genuine correlated hyperon–kaon pairs, $B'(\Dy)$ denotes the combinatorial background common to both distributions, and $B(\Dy)$ corresponds to the uncorrelated counterpart of $S(\Dy)$ generated through event mixing.

This transformation can be expressed as
\begin{align}
    S(x) &\Longrightarrow B(x) \, , \label{2509171509} \\
    B'(x) &\Longrightarrow B'(x) \, , \label{2509171510}
\end{align}
indicating that the combinatorial background remains unchanged under event mixing, while genuine correlations are converted into uncorrelated pairs.

Substituting Eqs.~(\ref{2509171532}) and (\ref{2509171533}) into Eq.~(\ref{2509191015}), the correlation function can be rewritten as
\begin{align}
    \CBS(\Dy) = c\left[S(\Dy) - B(\Dy)\right] \, . \label{2601101655}
\end{align}
This expression explicitly demonstrates that the combinatorial background $B'(\Dy)$ is removed by the subtraction procedure, and that $\CBS(\Dy)$ retains only the information associated with genuine hyperon–kaon correlations.

The schematic plots in Fig.~\ref{Same_Mix_Sub} illustrate that, during the transition from the uncorrelated background to the correlated signal, a fraction of hyperon–kaon pairs effectively “migrate” from regions with $\Dy < \Dy_1$ and $\Dy > \Dy_2$ into the interval $[\Dy_1, \Dy_2]$. 
This redistribution reflects an attractive correlation within this region. 
Consequently, the resulting $\CBS(\Dy)$ exhibits a shape similar to that shown in Fig.~\ref{Same_Mix_Sub_b}, indicating that kaons preferentially populate the relative-rapidity interval $\Dy \in [\Dy_1, \Dy_2]$. 

\subsection{Correlation strength and deviation}\label{Sec_EMDp}

To quantitatively characterize the shape of $\CBS(\Dy)$, which reflects the underlying correlation through the effective displacement of hyperon–kaon pairs from negative regions (e.g. area $A$ and $C$ in Fig.~\ref{Same_Mix_Sub}) into positive region (e.g. area $B$ in Fig.~\ref{Same_Mix_Sub}), we adopt a method analogous to the Energy Mover’s Distance (\EMD), or equivalently the one-dimensional Wasserstein distance~\cite{villaniOptimalTransportOld2009, duyExactStatisticalInference2023, rubnerEarthMoversDistance2000, komiskeHiddenGeometryParticle2020}.
We define
\begin{align}
    \text{WORK} &= \min_{\left\{f_{ij}\right\}} \sum_{i} \sum_{j} f_{ij} \left|\Dy_j - \Dy_i\right| \, , \label{2601101621} \\
    \text{EMD} &=  \frac{\sum_{i}\sum_{j}f_{ij}\left|\Dy_j-\Dy_i\right|}{\sum_{i}\sum_{j}f_{ij}}\,, \label{2601101638}
\end{align}
where indices $i$ and $j$ label bins of $\CBS(\Dy)$ in the negative region (e.g. $\Dy \in (-\infty,\Dy_1)\cup(\Dy_2,\infty)$ in Fig.\ref{Same_Mix_Sub}) and positive region (e.g. $\Dy \in [\Dy_1,\Dy_2]$ in Fig.\ref{Same_Mix_Sub}), respectively, and flow metric $f_{ij}$ represents the amount transported from bin $i$ to bin $j$ and satisfies the constraints
\begin{widetext}
    \begin{align}
        &f_{ij} \geq 0, \, \sum_{i}f_{ij} \leq \Theta\left(\CBS\left(\Dy_j\right)\right), 
        \sum_{j}f_{ij} \leq \Theta\left(-\CBS\left(\Dy_i\right)\right), \notag\\
        &\sum_{i}\sum_{j}f_{ij} = \sum_{\text{bins}}\Theta\left(\CBS\left(\Dy\right)\right) = \sum_{\text{bins}}\Theta\left(-\CBS\left(\Dy\right)\right) \,.\notag
    \end{align} 
\end{widetext}

The resulting $\EMD$ quantifies the average transport distance per unit weight under the optimal transport plan that minimizes the total WORK required to map the excess distributions in regions $A+C$ into region $B$, as illustrated in Fig.~\ref{Same_Mix_Sub_c}. 
In this sense, the $\EMD$ provides a quantitative measure of the difference between uncorrelated and correlated hyperon–kaon pair distributions, i.e. the correlation strength. 
A larger $\EMD$ value indicates that the units must be transported over greater distances from the negative to the positive region, implying a stronger correlation.
In this work, the $\EMD$ is computed using the open-source Python Optimal Transport (\POT) library~\cite{flamary2021pot, flamary2024pot}.

Based on the same optimal transport solution ${f_{ij}}$ obtained in Eq.~(\ref{2601101621}), we further define the Energy-Mover’s Displacement ($\EMDp$) as
\begin{align}
    \EMDp = \frac{\sum_{i}\sum_{j}f_{ij}\left(\Dy_j-\Dy_i\right)}{\sum_{i}\sum_{j}f_{ij}}\,.
\end{align}
In contrast to the $\EMD$ in Eq.~(\ref{2601101638}), which measures only the magnitude of the displacement, $\EMDp$ retains directional information. 
A positive (negative) $\EMDp$ indicates that correlated kaons are preferentially emitted at larger (smaller) relative rapidity $\Dy$. 
Therefore, when comparing two $\EMDp$ values, a larger (smaller) $\EMDp$ corresponds to faster (slower) correlated kaons.

The statistical uncertainties of $\EMDp$ are estimated using more than 5000 Monte Carlo samplings with Poissonian pseudo-experiments, where each bin in the hyperon–kaon distributions is assumed to be statistically independent. The resulting $\EMDp$ distributions are then fitted with a Gaussian function, and the corresponding $1\sigma$ width is taken as the statistical uncertainty. 

The uncertainty is influenced not only by the available statistics, but also by the intrinsic correlation strength.
In the absence of a genuine correlation—i.e., when the same-event distribution is identical to the mixed-event distribution—their subtraction is dominated purely by statistical fluctuations. 
In this case, the uncertainty does not decrease with increasing statistics. 

\section{Result and Discussion}\label{sec.3}
\subsection{Distribution component}

\begin{figure*}[!htb]
    \includegraphics[width=\hsize]{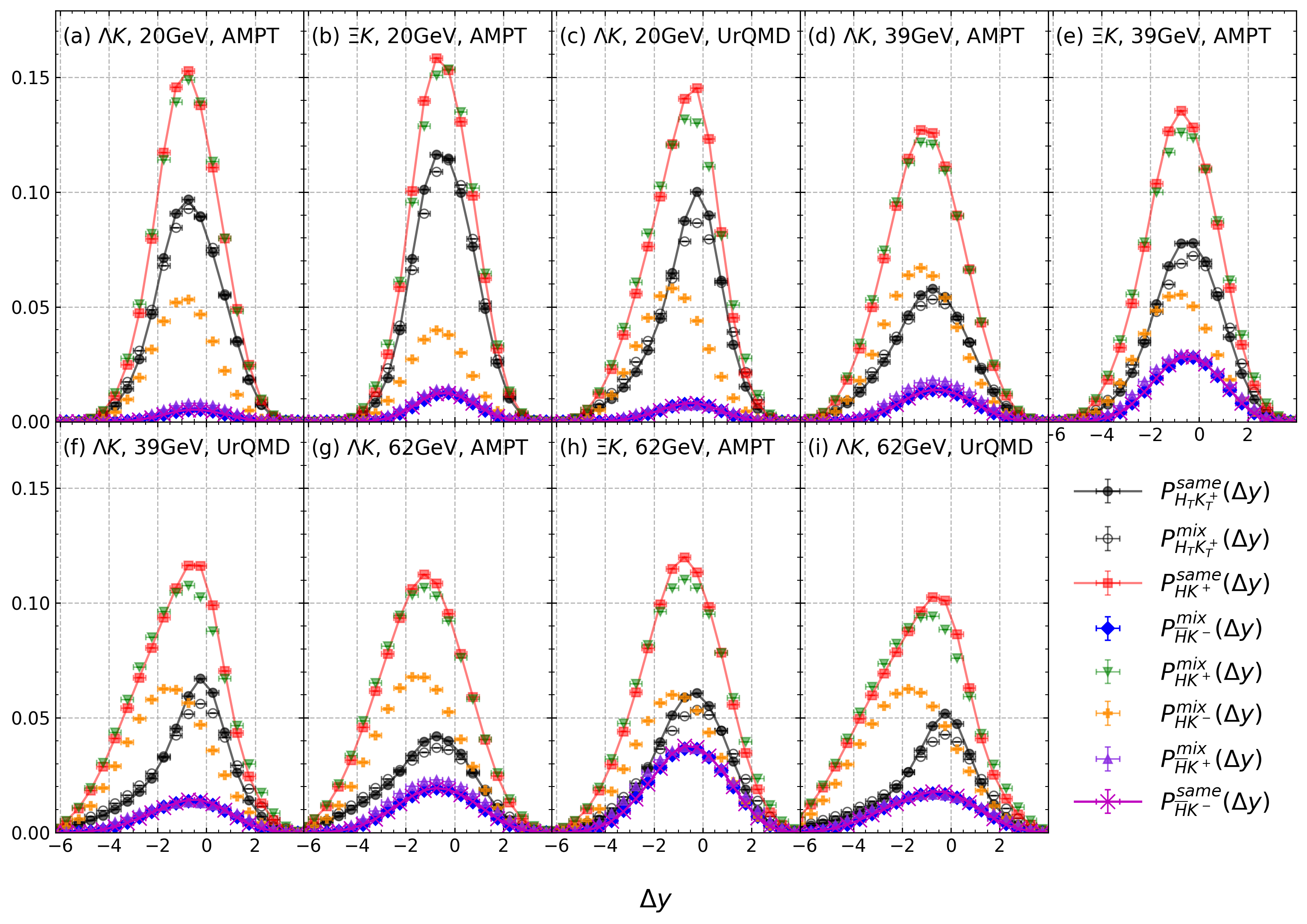}
    \caption{The \AMPT and \UrQMD results for $\Lambda K$ and $\Xi K$ pair distributions in \pAu collisions at $\sNN=20$, $39$, and $62$ GeV in full acceptance ($\left|\eta\right|<\infty$). 
    All distributions are normalized by the integral of $\sameHKp$ in each subfigure. 
    The same-event distributions are shown as data points with error bars connected by lines.}    
    \label{Distribution_Component_Full}
\end{figure*}

\begin{figure*}[!htb]
    \includegraphics[width=\hsize]{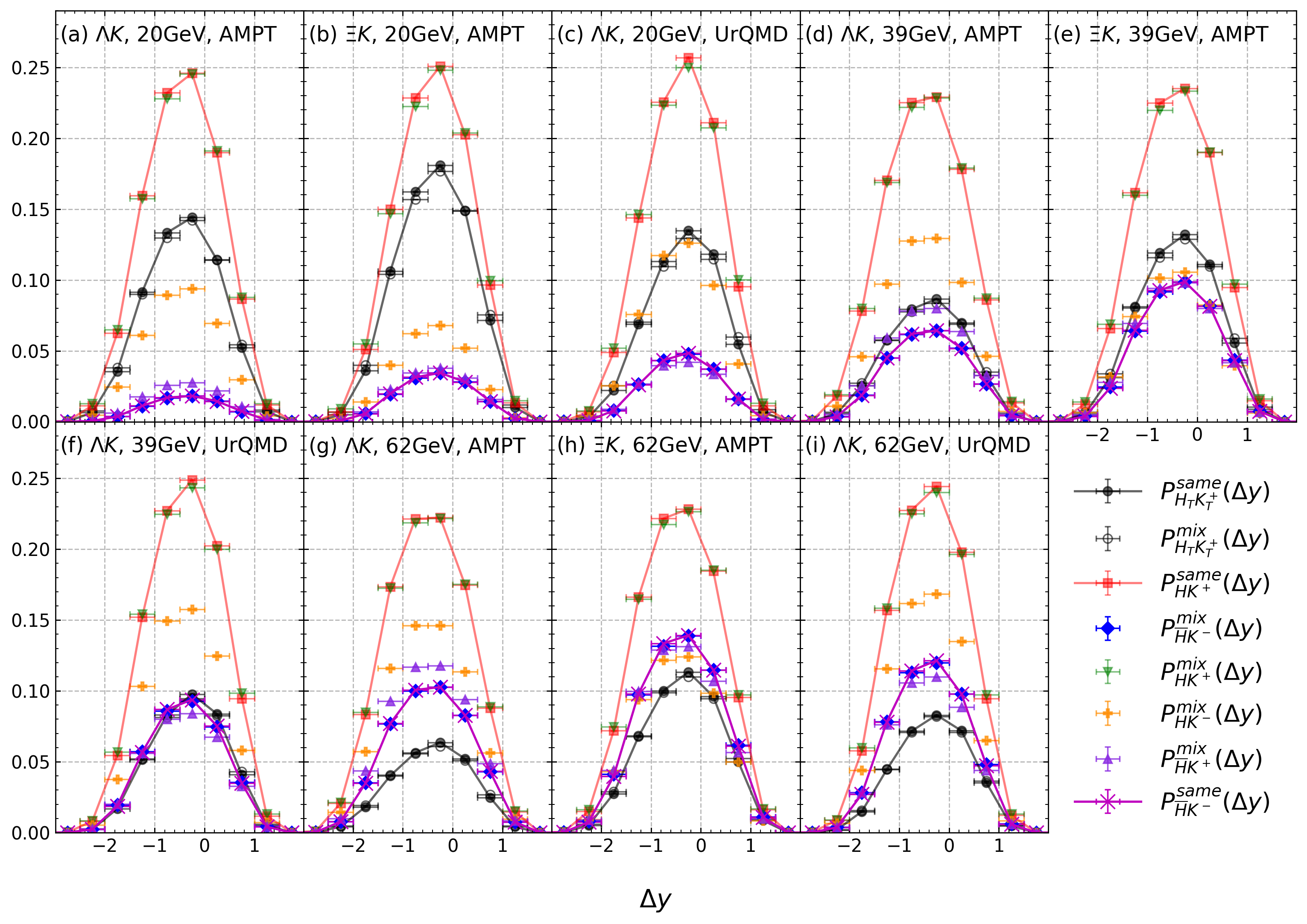}
    \caption{The \AMPT and \UrQMD results for $\Lambda K$ and $\Xi K$ pair distributions in \pAu collisions at $\sNN=20$, $39$, and $62$ GeV within detector acceptance ($\left|\eta\right|<1.5$).      All distributions are normalized by the integral of $\sameHKp$ in each subfigure.      The same-event distributions are shown as data points with error bars connected by lines.}
    \label{Distribution_Component}
\end{figure*}

The components of $\sameHTKpT$ and $\mixHTKpT$ in full acceptance are shown in Fig.~\ref{Distribution_Component_Full}. The red, black, and blue-violet curves in Fig.~\ref{Distribution_Component_Full} show $\sameHKp$ containing hyperon–kaon correlations from both scenario~1 and scenario~2, $\sameHTKpT$ containing contributions exclusively from scenario~1 and $\sameHbKm$ containing only contributions from scenario~2, respectively, illustrating the relative contributions of the different components.

By comparing panels (a), (d), and (g), panels (b), (e), and (h), as well as panels (c), (f), and (i) in Fig.~\ref{Distribution_Component_Full}, one observes that, with increasing collision energy, the relative contribution of $\sameHbKm$ with respect to $\sameHKp$ increases, whereas that of $\sameHTKpT$ decreases, for $\Lambda$ and $\Xi$ hyperons in \AMPT and for $\Lambda$ in \UrQMD. 
This trend indicates that a larger number of hyperon-kaon pairs originate from scenario~2 at higher collision energies. Such behavior reflects the increasing anti-hyperon-to-hyperon ratio discussed in Ref.~\cite{collaborationStrangeHadronProduction2020} and is consistent with expectations from baryon stopping effects.

Furthermore, by comparing panels (a) and (b), panels (d) and (e), as well as panels (g) and (h), one can see that both $P^{\text{same}}_{\Xi_T K^+_T}$ and $P^{\text{same}}_{\overline{\Xi} K^-}$ are larger than $P^{\text{same}}_{\Lambda_T K^+_T}$ and $P^{\text{same}}_{\overline{\Lambda} K^-}$, respectively. This observation indicates that both scenario~1 and scenario~2 contribute more significantly to $\Xi$ production than to $\Lambda$ production.

From Eq.~(\ref{2509072207}), this phenomenon can be attributed to the fact that the combinatorial background arising from $s$–$s$ and $\overline{s}$–$\overline{s}$ quark pairs (i.e., $\mixHbKm$ and $\mixHKp$) is larger for $\Lambda$ production than for $\Xi$ production. Since the ratio $\overline{\Xi}/\Xi$ is smaller than $\overline{\Lambda}/\Lambda$, fewer anti-hyperons and $K^-$ mesons are produced in the $\Xi$ channel, leading to reduced contributions from $\mixHbKm$ and $\mixHKp$.

Consequently, in full-acceptance measurements of the hyperon–$K^+$ pair distributions in \pAu collision system, hyperons with higher strangeness content are expected to exhibit a larger fraction of genuine contributions from both scenario~1 ($\sameHTKpT$) and scenario 2 ($\sameHPKpP$), and a correspondingly smaller (larger) contribution from combinatorial backgrounds associated with $s$–$s$ ($\overline{s}$–$\overline{s}$) quark pairs, i.e. $\mixHKm$ ($\mixHbKp$).

The components of $\sameHTKpT$ and $\mixHTKpT$ within the detector acceptance $\lvert \eta \rvert < 1.5$ are shown in Fig.~\ref{Distribution_Component}. Overall, the relative contributions and qualitative trends observed in full acceptance are preserved under the detector acceptance. However, for $\Lambda$ hyperons, the differences between \AMPT and \UrQMD become more pronounced when the detector acceptance is applied. As shown in panels (a) and (c), (d) and (f), and (g) and (i) of Fig.~\ref{Distribution_Component_Full} and Fig.~\ref{Distribution_Component}, $\sameHbKp$ is systematically larger in \UrQMD than in \AMPT within the detector acceptance, whereas the two models yield nearly identical results in full acceptance. These results indicate a difference between the two models in the hyperon yields originating from scenario~2 in the mid-rapidity region. 

\subsection{Correlation function}
\subsubsection{$\CBS$}

\begin{figure*}[!htb]
    \includegraphics[width=\hsize]{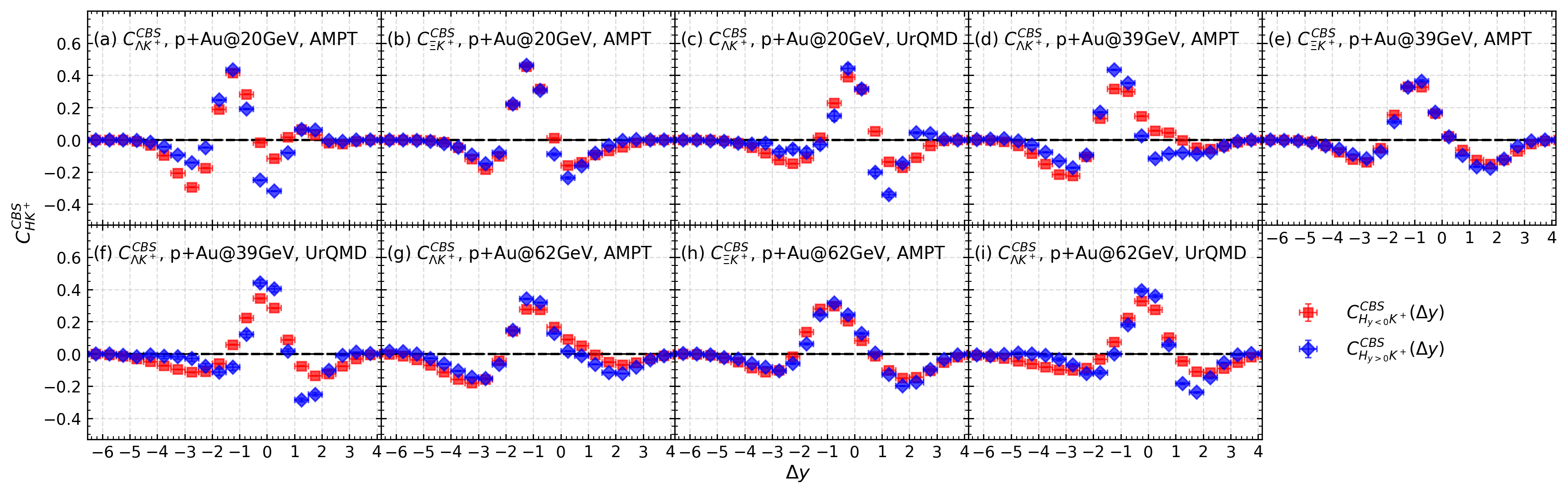}
    \caption{The \AMPT and \UrQMD result for $\CBSLambda(\Dy)$ and $\CBSXi(\Dy)$ in \pAu collisions at $\sNN=20$, $39$ and $62$ GeV in full acceptance ($\left|\eta\right|<\infty$). 
    The subscript $y > 0$ denotes hyperons with positive rapidity, corresponding to the initial proton-going direction, while $y < 0$ indicates hyperons with negative rapidity, corresponding to the Au-going direction. }
    \label{FIG_dRap_List_Full}
\end{figure*}

\begin{figure*}[!htb]
    \includegraphics[width=\hsize]{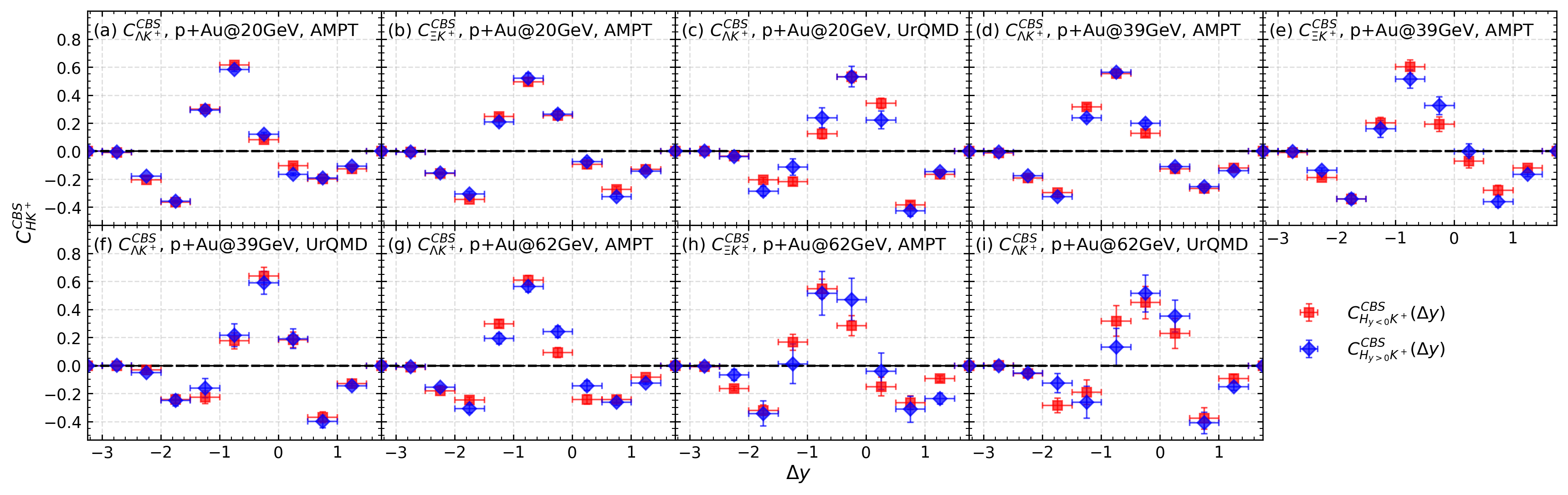}
    \caption{The \AMPT and \UrQMD results for $\CBSLambda(\Dy)$ and $\CBSXi(\Dy)$ in \pAu collisions at $\sNN=20$, $39$ and $62$ GeV within detector acceptance ($\left|\eta\right|<1.5$). Other notations are the same as in Fig.~\ref{FIG_dRap_List_Full}.}
    \label{FIG_dRap_List}
\end{figure*}

The correlation functions $\CBS$ for the full and detector acceptances are shown in Fig.~\ref{FIG_dRap_List_Full} and Fig.~\ref{FIG_dRap_List}, respectively. All sub-figures exhibit trends similar to the schematic plots in Fig.~\ref{Same_Mix_Sub_b}.
Based on the discussion in Sec.~\ref{Sec_CBS}, where the shape of $C^{\text{CBS}}_{H K^+}(\Dy)$ reflects the preferred relative rapidity of $K^+$ mesons correlated with hyperons, both $C^{\text{CBS}}_{\Lambda K^+}(\Dy)$ and $C^{\text{CBS}}_{\Xi K^+}(\Dy)$ in the \AMPT model, as well as $C^{\text{CBS}}_{\Lambda K^+}(\Dy)$ in the \UrQMD model at all collision energies, indicate that kaons correlated with hyperons preferentially populate a relative rapidity range of approximately $\Dy \in (-2,1)$ for full acceptance and $\Dy \in (-1.5,0.5)$ for detector acceptance. 
By comparing the shapes of the correlation functions, one can determine whether the correlated kaons are preferentially slower or faster in relative rapidity $\Dy$ for different hyperon species, collision energies, and models. 
These differences are quantified by the corresponding $\EMDp$ and $\EMD$ values, as shown in Fig.~\ref{FIG_dDy_sNN_Full} and Fig.~\ref{FIG_dDy_sNN}.

\subsubsection{$\EMD$ and $\EMDp$}

\begin{figure}
    \includegraphics[width=\hsize]{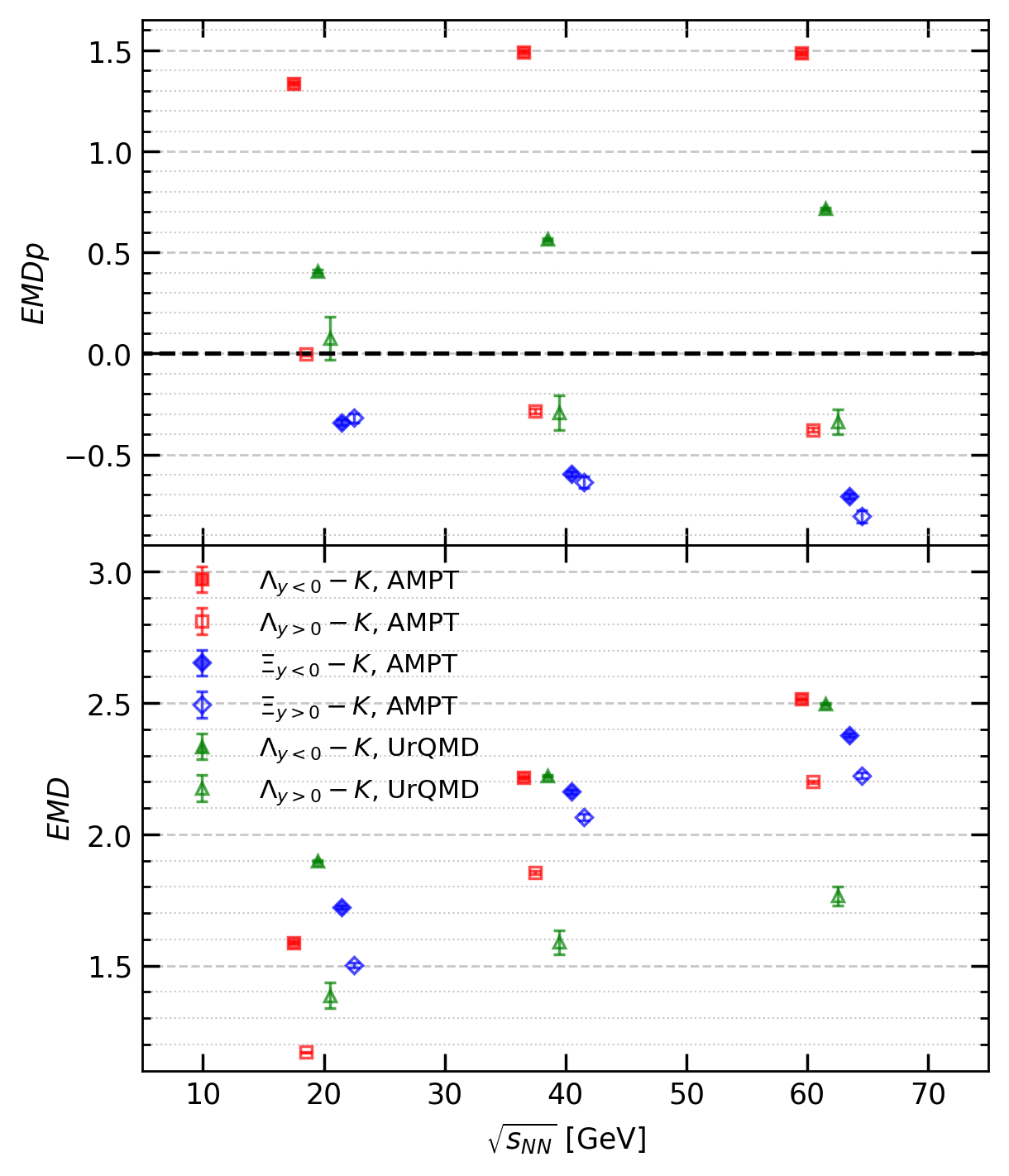}
    \caption{The \AMPT and \UrQMD results for the $\EMDp$ and $\EMD$ in $C^{\text{CBS}}_{\Lambda K^+}(\Dy)$, $C^{\text{CBS}}_{\Xi K^+}(\Dy)$ in \pAu collisions at $\sNN=20$, $39$ and $62$ GeV in full acceptance ($\left|\eta\right|<\infty$). 
    The subscripts $y>0$ and $y<0$ have the same meaning as in Fig.~\ref{FIG_dRap_List}. For better visibility, data points at identical energies are slightly shifted along the horizontal axis. }
    \label{FIG_dDy_sNN_Full}
\end{figure}

\begin{figure}
    \includegraphics[width=\hsize]{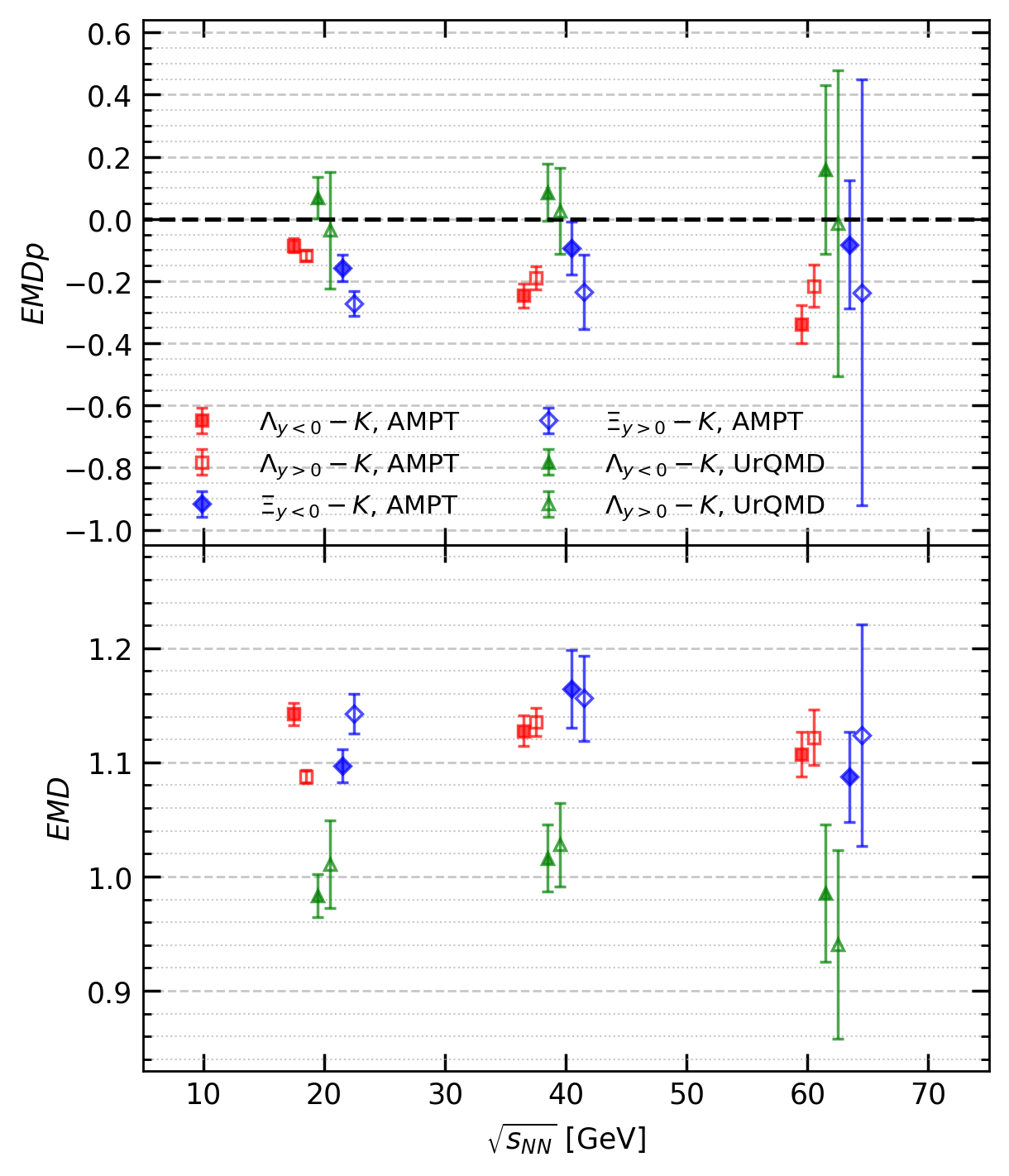}
    \caption{The \AMPT and \UrQMD results for the $\EMDp$ and $\EMD$ in $C^{\text{CBS}}_{\Lambda K^+}(\Dy)$, $C^{\text{CBS}}_{\Xi K^+}(\Dy)$ in \pAu collisions at $\sNN=20$, $39$ and $62$ GeV in detector acceptance ($\left|\eta\right|<1.5$). 
    Other notations are the same as in Fig.~\ref{FIG_dDy_sNN_Full}}. 
    \label{FIG_dDy_sNN}
\end{figure}

Differences between the correlation functions for hyperons with positive and negative rapidity, $\CBSyp\left(\Dy\right)$ and $\CBSym\left(\Dy\right)$, are observed in Fig.~\ref{FIG_dRap_List_Full} for full acceptance and Fig.~\ref{FIG_dRap_List} for detector acceptance ($\lvert \eta \rvert < 1.5$). The simulation results suggest that hyperons carrying baryon number transported from the initial proton and the \Au nucleus correlate with kaons with systematically different strengths, leading to the differences which can be seen by comparing the solid and hollow markers in Fig.~\ref{FIG_dDy_sNN_Full} and Fig.~\ref{FIG_dDy_sNN}. 

For the full acceptance case, the difference between $\CBSLambdayp\left(\Dy\right)$ and $\CBSLambdaym\left(\Dy\right)$ is pronounced in both models, whereas the difference between $\CBSXiyp\left(\Dy\right)$ and $\CBSXiym\left(\Dy\right)$ is much smaller. 
At $\sqrt{s_{NN}} = 20$ and $39$ GeV, the corresponding $\EMDp$ values for $\CBSXiyp\left(\Dy\right)$ and $\CBSXiym\left(\Dy\right)$ are nearly identical, while at $\sqrt{s_{NN}} = 62$ GeV a slight separation can be observed.  
This may be attributed to the fact that $\Lambda$ hyperons are more likely to retain a larger valence-quark component from incident nucleons than $\Xi$ hyperons and more strongly influenced by subsequent collisions, implying that the incident proton undergoes more collisions than nucleons in \Au nuclei before fragmenting into $\Lambda_{y>0}$ ($\Xi_{y>0}$), which consequently carries lower energy and is more likely to be separated from slower kaons than $\Lambda_{y<0}$ ($\Xi_{y<0}$) during the BNT. This mechanism may also account for the stronger correlation observed for $\CBSym\left(\Dy\right)$ compared with $\CBSyp\left(\Dy\right)$, as reflected by the $\EMD$ values shown in Fig.~\ref{FIG_dDy_sNN_Full}. Specifically, $\CBSym\left(\Dy\right)$ (solid points) exhibits systematically larger $\EMD$ values than $\CBSyp\left(\Dy\right)$ (hollow points) for both hyperon species and for both models.

Within the detector acceptance, however, the differences between $\CBSym(\Dy)$ and $\CBSyp(\Dy)$ become significantly reduced, particularly for $\CBSLambda(\Dy)$ in both \AMPT and \UrQMD. As shown in Fig.~\ref{FIG_dDy_sNN}, the distributions of $\CBSLambdaym(\Dy)$ and $\CBSLambdayp(\Dy)$ in both models, as well as $\CBSXiym(\Dy)$ and $\CBSXiyp(\Dy)$ in \AMPT, are very close to each other. 
This reduction may be understood within the valence-quark picture, where hyperon–kaon correlations become weaker in the mid-rapidity region, consistent with the expected suppression of BNT in this framework. If baryon junction interactions are present, one would anticipate a larger separation between $\CBSym(\Dy)$ and $\CBSyp(\Dy)$ than that shown in Fig.~\ref{FIG_dDy_sNN} in experimental data. 

\subsubsection{Energy dependence}

As shown in Fig.~\ref{FIG_dDy_sNN_Full}, in full acceptance, a common energy dependence is observed for $\CBSLambdayp(\Dy)$ and $\CBSXiyp(\Dy)$ in both cases, the $\EMDp$ values decrease consistently with increasing collision energy.
This behavior indicates that hyperons carrying baryon number from the incident proton tend to be correlated with relatively slower kaons at higher energies.
In contrast, within a baryon junction scenario, an opposite trend may be anticipated. 
If the gluonic field carries the baryon number and experiences stronger stopping at higher energies, while the valence quarks forming kaons lose relatively less energy, a larger rapidity separation between hyperons and kaons would be expected, corresponding to larger $\EMDp$ values. 

The $\EMD$ results shown by hollow and solid points in Fig.~\ref{FIG_dDy_sNN_Full} exhibit a monotonic increasing trend with collision energy $\sNN$ for both $\Lambda$ and $\Xi$ hyperons in both models, indicating a stronger hyperon-kaon correlation at higher collision energies. The results suggest that, in higher-energy collisions, the incident nucleons tend to fragment into more strongly interacted hyperons and kaons.

\subsubsection{$\Lambda$ vs. $\Xi$}

In \AMPT, the $\EMDp$ values for $\CBSXiyp(\Dy)$ are systematically smaller than those for $\CBSLambdayp(\Dy)$ at all energies in full acceptance. 
This may be due to the stronger stopping experienced by multi-strange hyperons during BNT from incident nucleons. Consequently, they tend to correlate with slower kaons in the valence quark picture, consistent with the energy dependence discussed above.
This interpretation is also consistent with the picture that transferring a larger number of valence quarks from the proton to kaons during the fragmentation process enhances the hyperon-kaon correlation strength, as may be inferred from the $\EMD$ data points in Fig.~\ref{FIG_dDy_sNN_Full}, where, $\CBSXiyp(\Dy)$ tends to be larger than $\CBSLambdayp(\Dy)$ at low collision energies (20 and 39 GeV), and close to each other at higher energy (62 GeV).

In brief, stronger baryon stopping, for example at higher collision energies or in baryon number transport to hyperons with greater valence strangeness, leads to a slower baryon number carrier.
The results from \AMPT and \UrQMD, based on the valence quark picture, show slower kaons as expected. In contrast, an opposite tendency, namely slower hyperons, is anticipated in the baryon junction mechanism.

\subsubsection{Model dependence}

Although the maximum of $\CBSLambdayp(\Dy)$ in \UrQMD is shifted toward larger $\Dy$ values compared to \AMPT within full acceptance, around $\Dy \approx 1$, the difference in the corresponding $\EMDp$ values remains smaller than the associated uncertainties as showed in Fig.~\ref{FIG_dDy_sNN_Full}, indicating consistency between the two models.
In contrast, the $\EMDp$ of $\CBSLambdaym(\Dy)$ shows a significant discrepancy between \AMPT and \UrQMD. A direct reason is that $\CBSLambdaym(\Dy)$ in \UrQMD is systematically larger than in \AMPT in the positive-$\Dy$ region. Consequently, more kaons are effectively shifted toward larger $\Dy$ values in \UrQMD. This difference likely originates from the distinct implementations of BNT during sequential collisions in the two models. 

For the $\EMD$ values, however, $\CBSLambdayp(\Dy)$ tends to be closer to $\CBSLambdaym(\Dy)$ at higher collision energies ($\sNN=39$ and $62$ GeV). In addition, $\CBSLambdayp(\Dy)$ exhibits a gentler energy dependence than $\CBSLambdaym(\Dy)$ in \UrQMD\ compared with \AMPT, indicating a clear energy dependence of the correlation strength.

The discussion above of $\CBSLambdayp(\Dy)$ and $\CBSXiyp(\Dy)$ in Fig.~\ref{FIG_dDy_sNN_Full} indicates that, in full acceptance, the correlation between hyperons at positive rapidity (the proton-going direction) and kaons provides a more reliable probe for observing BNT in \pAu collisions than correlations involving hyperons at negative rapidity (the Au-going direction). 
The proton-going side is therefore expected to offer cleaner dynamics and a more model-independent baseline. 

\section{Summary}

Hyperon-kaon correlations provide a sensitive probe of BNT and baryon fragmentation in high-energy collisions. 
The $\CBS(\Dy)$ correlation functions defined in Eq.~\ref{2509191015} characterize the preferred relative rapidity $\Dy$ between correlated kaons and hyperons during the transport of baryon number from the incident nucleon to strange baryons. 

By analyzing $\CBS(\Dy)$ and the corresponding $\EMDp$ values in \pAu\ collisions at $\sqrt{s_{NN}}=20$, $39$, and $62$~GeV using the \AMPT\ and \UrQMD\ models, we investigate the rapidity separation between the baryon number carrier and the transferred valence quarks in hyperon production. Correlations between forward (positive-rapidity) hyperons—emitted along the incident proton direction—and kaons provide a clearer indication of proton fragmentation. We find that hyperons emitted in the proton-going direction provide a prospective probe for understanding the BNT effect in p+A collisions.

We further compare the correlation patterns for different hyperons($\Lambda$ and $\Xi$), examining their energy and model dependence. In addition to results obtained in full acceptance, we evaluate the impact of experimental detector acceptance constraints. 

Our study introduces a novel approach to probing baryon fragmentation within the BNT framework, establishing a baseline for future experimental analyses aimed at testing the presence of baryon junction interactions. In the absence of such a mechanism, this baseline offers a clear reference prediction, which future experiments with large rapidity coverage can test to further advance our understanding of the BNT effect.


\bibliographystyle{nst}
\bibliography{main}       

\end{document}